\documentclass[prl,twocolumn,showpacs,preprintnumbers]{revtex4}

\usepackage[dvipdfmx]{graphicx}
\usepackage{amsmath}
\usepackage{amssymb}
\usepackage{times}
\usepackage{color}
\usepackage{ulem}

\def\D{\Delta}

\def\e{\epsilon}

\def\h{\eta}

\def\r{\rho}
\def\S{\Sigma}
\def\s{\sigma}

\def\w{\omega}
\def\ua{\uparrow}
\def\da{\downarrow}

\begin{document}

\title{Hidden fermionic excitation in the superconductivity of the strongly attractive Hubbard model}

\author{Shiro Sakai$^1$, Marcello Civelli$^2$, Yusuke Nomura$^3$, and Masatoshi Imada$^4$}

\affiliation{$^1$RIKEN Center for Emergent Matter Science, 2-1 Hirosawa, Wako, Saitama 351-0198, Japan\\
$^2$Laboratoire de Physique des Solides, Universit\'e Paris-Sud, CNRS, UMR 8502, F-91405 Orsay Cedex,
France\\
$^3$Centre de Physique Th\'eorique, \'Ecole Polytechnique, CNRS UMR 7644, 91128 Palaiseau, France\\
$^4$Department of Applied Physics, University of Tokyo, 7-3-1 Hongo, Tokyo 113-8656, Japan}
\date{\today}

\begin{abstract}
We scrutinize the real-frequency structure of the self-energy in the superconducting state of the attractive Hubbard model within the dynamical mean-field theory.
Within the strong-coupling superconducting phase which has been understood in terms of the Bose-Einstein condensation in the literature, we find two qualitatively different regions crossing over each other. In one region close to zero temperature, the self-energy depends on the frequency only weakly at low energy. On the other hand, in the region close to the critical temperature, the self-energy shows a pole structure. 
The latter region becomes more dominant as the interaction becomes stronger.
We reveal that the self-energy pole in the latter region is generated by a coupling to a 
hidden {\it fermionic} excitation.
The hidden fermion persists in the normal state, where it yields a pseudogap.
We compare these properties with those of the repulsive Hubbard model relevant for high-temperature cuprate superconductors, showing that hidden fermions are a key common ingredient in strongly correlated superconductivity.
\end{abstract}
\maketitle

%---introduction
A range of metals show superconductivity below a critical temperature ($T_\text{c}$), at which paired electrons acquire a spatial coherence.
In conventional superconductors, the pairing mechanism is well described by the Bardeen-Cooper-Schrieffer (BCS) theory \cite{bcs57}.
However, the BCS theory does not straightforwardly apply when the attractive interaction between electrons is strong. In this case, the electron pairing occurs at a temperature higher than $T_\text{c}$ and superconductivity arises when the electron pairs (regarded as composite bosons) go through the Bose-Einstein condensation (BEC) at a lower temperature \cite{eagles69}.
Such superconductivity in the BEC regime has been explored for ultracold fermionic atom systems \cite{jochim03,zwierlein03,bloch08}. In fact, tightly-bound pairs \cite{greiner03} and an associated pseudogap behavior \cite{gaebler10,feld11} have been observed for $^{40}$K gas in the strongly-interacting region.
A recent experiment also suggested that the superconductivity in FeSe is in the BCS-BEC  crossover regime \cite{kasahara14}.
The preformed pairing has also been proposed in the context of cuprate high-$T_\text{c}$ superconductors \cite{anderson87,emery95}. Although it looks unlikely that the preformed pairing solely can explain the whole pseudogap behavior in the cuprates \cite{tacon06,tanaka06,kondo07,sakai13}, it may be relevant in a region close to $T_\text{c}$ around the optimal doping \cite{xu00,wang01}.

On the theoretical side, the crossover from BCS to BEC \cite{legett80,nozieres85,micnas90,bloch08} has been intensively studied with continuous \cite{haussmann93,tokumitu93,pieri04,kinnunen04,ohashi05,tsuchiya09,watanabe10} or lattice \cite{robas81,randeria92,rodero92,santos94,trivedi95,singer96,kyung01,sewer02,
paiva04,dupuis04,burovski06,bulgac06,zhao06,tamaki08} fermion models, both of which give a similar phase diagram.
The latter is, however, more tractable with numerical simulations and allows us to employ nonperturbative methods to study this problem.
In particular, the dynamical mean-field theory (DMFT) \cite{georges96}, which becomes exact in infinite spatial dimensions, is a suitable tool to study dynamical properties of the lattice models. In fact, the DMFT and its extensions have been extensively applied to the attractive Hubbard model \cite{freericks94,keller01,keller02,capone02,toschi05,toschi05-2,laloux,kyung06,garg05,bauer09,bauer09-2,su10,koga11,murakami13,staar14,peters15}.
Among the results, for strong coupling, a pseudogap in the spectral function has been found above $T_\text{c}$ \cite{kyung06,bauer09,su10,koga11,peters15}.
Further, the pseudogap state was found to be separated from a Fermi-liquid metal at weak coupling by a first-order pairing transition \cite{keller01,keller02,capone02,laloux}.
In light of an electron-hole transformation \cite{micnas90}, this 
first-order transition is mapped onto a well-known Mott metal-insulator transition in the repulsive half-filled Hubbard model under an external magnetic field. 
Since the Mott gap in the repulsive model is characterized by an emergent pole in the self-energy, a self-energy pole emerges equivalently in the pseudogap state of the attractive model  \cite{pieri04,bauer09,peters15}, signaling the formation of the preformed pairs above $T_\text{c}$.

It is a nontrivial open problem to determine the fate of the self-energy pole below $T_\text{c}$ since the sudden appearance of the pole in the normal phase appears incompatible with a smooth crossover of superconducting properties from weak to strong coupling. 
At weak coupling in the BCS region, the dynamics of quasiparticles, described by the low-frequency dependence of the self-energies, is well understood. 
On the other hand, at strong coupling in the BEC region, the superconducting transition is well described by the boson condensation of tightly-bound Cooper pairs \cite{micnas90}, while the dynamics of the original fermions, which is directly relevant to the single-particle spectroscopies such as the radiofrequency \cite{stewart08}/photoemission spectroscopy, has not been fully addressed.

In this paper we propose an alternative (``fermionic") viewpoint to the BEC of tightly-bound Cooper pairs.
This decription also applies to the intermediate-coupling region where the bosonic picture does not hold anymore, bridging the BEC limit with the well known BCS region.
Our main purpose is to reveal that the low-energy quasiparticle dynamics in the BEC region is governed by a coupling to a {\it fermionic} excitation (hidden fermion) arising from the strong interaction. The hidden fermion generates the self-energy pole in the normal states and persists below $T_\text{c}$,
similarly to what has been found in the superconducting state of the repulsive Hubbard model \cite{sakai14}.
As the temperature $T$ lowers further, however, the hidden fermion loses the weight and eventually vanishes, in marked contrast with the repulsive case. 
These results set out for a unified view of the superconducting transition of tightly-bound Cooper pairs and its relation with the Mott physics.

%---method

We consider the normal and superconducting states of the infinite-dimensional Hubbard model on a Bethe lattice
\begin{align}
H=-t\sum_{\langle ij \rangle\s} a_{i\s}^\dagger a_{j\s}-\mu\sum_i n_{i\s} +U\sum_i n_{i\ua}n_{i\da},
\label{HM}
\end{align}
which has the advantage to be exactly solvable with the DMFT \cite{georges96}.
Here the onsite interaction $U <0$ is attractive, $t$ is the hopping integral and $\mu$ is the chemical potential. $a_{i\s}^\dagger (a_{i\s})$ creates (annihilates) a fermion with spin $\s$ at site $i$, and $n_{i\s}\equiv a_{i\s}^\dagger a_{i\s}$. The bare density of states is $\r_0(\w)=\frac{2}{\pi D^2}\sqrt{D^2-\w^2}$ with $D=2$, which determines the energy scale in the paper. 
In order to avoid possible peculiarities at half filling, we set the fermion density per site to be $n=0.8$ \cite{note1}.

The DMFT maps this model onto an impurity problem, which we solve with a finite-$T$ extension of the exact diagonalization (ED) method \cite{liebsch12} elaborated in Supplementary Information.
The ED solver allows us to study precise real-frequency ($\w$) structures of the self-energies, which are of our main interest.
For superconductivity, we calculate the normal and anomalous self-energies, $\S^\text{nor}$ and $\S^\text{ano}$, related to the normal Green's function by 
\begin{align}
G(\e,\w)&=\left[\w -\e+\mu-\S^\text{nor}(\w) -W(\e,\w)\right]^{-1}\label{G}
\end{align}
with
\begin{align}
W(\e,\w)&\equiv \frac{ \S^\text{ano}(\w)^2 }{\w+\e-\mu+\S^\text{nor}(-\w)^\ast}. \label{W}
\end{align}
Since in DMFT $G$ depends on particle's momentum only through the bare-dispersion energy $\e$, the density of states (DOS) is calculated from Eq.~(\ref{G}) as 
$-\frac{1}{\pi}\text{Im}\int_{-D}^D d\e \r_0(\e) G(\e,\w)$.

\begin{figure}[tb]
\center{
\includegraphics[width=0.48\textwidth]{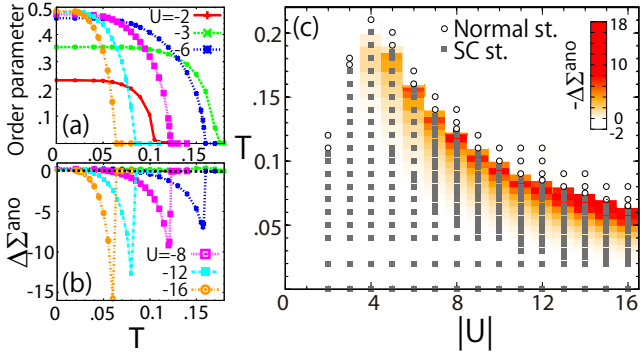}}
\caption{ (Color online) Temperature dependence of (a) the superconducting order parameter and (b) $\varDelta\S^\text{ano}\equiv \S^\text{ano}(0)-\S^\text{ano}(\infty)$, for various values of $U$. (c) Phase diagram against $T$ and $|U|$, with intensity plot of $-\varDelta\S^\text{ano}$.}
\label{pd}
\end{figure}

%---result
Figure \ref{pd}(a) plots the superconducting order parameter, $\langle a_{i\ua}a_{i\da}\rangle$, against $T$ for various values of $U$. 
We define $T_\text{c}$ as the lowest temperature above which $\langle a_{i\ua}a_{i\da}\rangle$ is less than 0.01. 
Thus-determined phase boundary in Fig.~\ref{pd}(c) is consistent with previous works \cite{toschi05,toschi05-2,bauer09,bauer09-2,koga11,peters15}.

\begin{figure}[tb]
\center{
\includegraphics[width=0.48\textwidth]{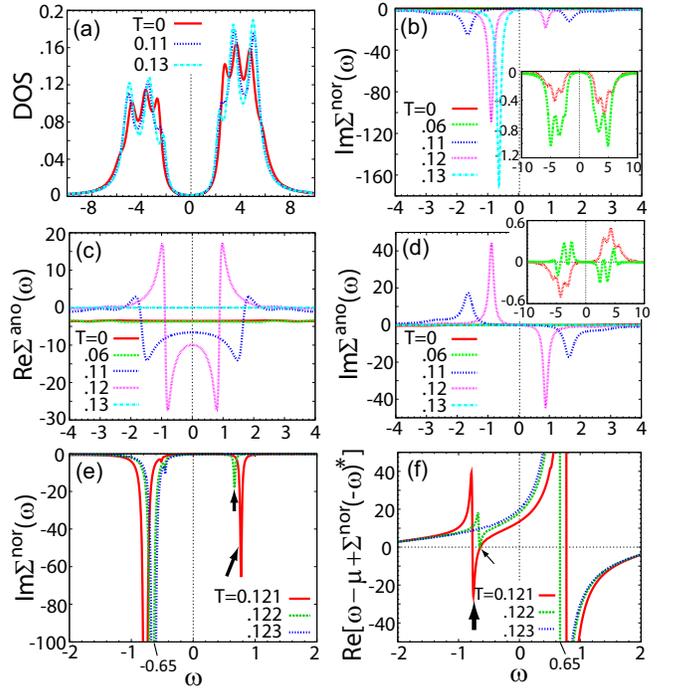}}
\caption{(Color online) (a-d) Various real-frequency properties at $U=-8$ in the normal  ($T=0.13 > T_\text{c} \simeq 0.123$) and superconducting ($T=0.12, 0.11, 0.06, 0$) states.
The insets to panels (b) and (d) are the enlarged views for $T=0$ and 0.06.
(e), (f) Change of Im$\S^\text{nor}$ and Re$[\w-\mu+\S^\text{nor}(-\w)^*]$ across $T_\text{c}$, where $\h_0=0.01$ (see Supplementary Information) is used to sharpen the pole and zero features. The arrows in panel (e) indicate the pole created below $T_\text{c}$. The thick and thin arrows in panel (f) indicate the pole and zero of the plotted quantity, respectively, for $T=0.121$, which are created in the superconducting state.} 
\label{dos}
\end{figure}

For $|U|<4$, $T_\text{c}$ increases with $|U|$. 
In this region, the superconductivity follows well the BCS theory; a gap opens in the DOS only below $T_\text{c}$; for  $T > T_\text{c}$ $\S^\text{nor}$ shows a Fermi-liquid-like behavior, i.e., Im$\S^\text{nor}\propto \w^2$ (not shown here). 

On the other hand, for $|U|>4$, in the so-called BEC region, $T_\text{c}$ decreases with $|U|$. In this case the normal state displays preformed pairs and a gap (called pseudogap) in the DOS even above $T_\text{c}$ \cite{kyung06,bauer09,su10,koga11,peters15}, as shown in Fig.~\ref{dos}(a) for $U=-8$ and $T=0.13>T_\text{c}\sim0.123$.
This gap results from a (thermally-smeared) pole in $\S^\text{nor}$ (because $W=0$), which can be seen for instance at $\w\simeq -0.7$ for $T=0.13$ in Fig.~\ref{dos}(b). 
The first finding in this study is that this pole survives even below $T_\text{c}$, symmetrically split with respect to $\omega=0$ by the appearance of superconductivity which creates Bogoliubov electron-hole branches. These peaks can be for example seen for $T=0.12$ at $\w\simeq \pm 0.9$ in  Fig.~\ref{dos}(b).
The second remarkable finding is that $\S^\text{ano}(\w)$ also shows the appearance of these poles at the {\it same} energies $\w=\pm 0.9$, which give considerable dynamical structure to Re$\S^\text{ano}$ [Fig.~\ref{dos}(c)] and Im$\S^\text{ano}$ [Fig.~\ref{dos}(d)]. Surprisingly, the weight of these poles diminish as $T$ lowers further [Fig.~\ref{dos}(d)], becoming indiscernible for $T<0.06$ (inset). Consequently, Re$\S^\text{ano}$ is nearly flat at low temperatures [red and green curves in the panel (c)]. The value agrees with what is expected for a static mean-field decoupling of the interaction term, $U n_{i\ua}n_{i\da}$, in Eq.~(\ref{HM}); its anomalous channel is given by $U(\langle a_{i\da}^\dagger a^\dagger_{i\ua} \rangle  a_{i\ua} a_{i\da}+  \langle a_{i\ua} a_{i\da} \rangle  a^\dagger_{i\da} a^\dagger_{i\ua})$.

\begin{figure}[tb]
\center{
\includegraphics[width=0.48\textwidth]{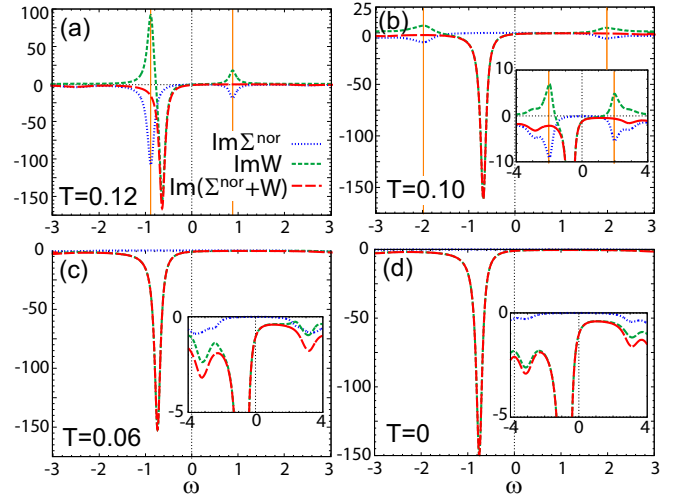}}
\caption{(Color online) Presence/absence of the cancellation of the peaks between Im$\S^\text{nor}$ and Im$W$ for various temperatures at $U=-8$.  The insets are the enlarged views of the each panel.
Orange vertical lines indicate the energy of the peaks which cancel out.}
\label{cancel}
\end{figure}

Despite this drastic evolution of the self-energies (i.e., appearance of the poles in $\S^\text{ano}$ at $T_\text{c}$ and their subsequent disappearance at lower $T$), the gap in the DOS does not substantially change with lowering $T$ [Fig.~\ref{dos}(a)] \cite{note3}.
The singularity generating the spectral gap is in fact transformed from the normal self-energy poles for $T>T_\text{c}$ to a pole in $W$, i.e. a zero of Re$[\w+\e-\mu+\S^\text{nor}(-\w)^*]$, for $T<T_\text{c}$.
In order to elucidate this point, we plot in Figs.~\ref{dos}(e) and (f), Im$\S^\text{nor}$ and Re$[\w-\mu+\S^\text{nor}(-\w)^*]$, respectively, in the vicinity of $T_\text{c}$ \cite{note5,note6}. (Compare also the temperature dependences of the peak weights between Im$\S^\text{nor}$ and Im$W$ in Fig.~\ref{cancel}.)
At $T=0.123\gtrsim T_\text{c}$, Im$\S^\text{nor}$ shows a single peak at $\w=\w_\text{nor}=-0.65$, while, as mentioned above, at $T < T_\text{c}$ another pole in $\S^\text{nor}$ is created at the particle-hole symmetric energy, $\w=-\w_\text{nor}$, as indicated by the arrows in Fig.~\ref{dos}(e).
The corresponding pole in Re$[\w-\mu+\S^\text{nor}(-\w)^*]$ (arising from the pole of $\S^\text{nor}(-\w)^*$) is indicated by a thick arrow for $T=0.121$ in panel (f).
This new pole is accompanied by a new zero of Re$[\w-\mu+\S^\text{nor}(-\w)^*]$, as indicated by the thin arrow in the same panel.
This new zero is the origin of the strong negative peak in Im$W$ and Im$(\S^\text{nor}+W)$ [Fig.~\ref{cancel}(a)], which yields the spectral gap in the superconducting state. 
This is similar to the BCS superconductors in that a zero of Re$[\w+\e-\mu+\S^\text{nor}(-\w)^*]$ gives the superconducting gap.
However, as discussed above, the zero in our case is tightly connected to the presence of the normal self-energy poles which are a smooth continuation of those already present in the normal state.
Thus, the pseudogap and superconducting gap are closely related with each other while they involve different singularities.

%---cancellation

Another notable finding is that the pole of $\S^\text{nor}$ and one of the poles of $W$ cancel with each other in $G$ [Eq.~(\ref{G})].
To see this property, we plot in Fig.~\ref{cancel} Im$\S^\text{nor}$, Im$W$ and their sum  Im$(\S^\text{nor}+W)$ at $\e=0$ for several temperatures at $U=-8$. 
Surprisingly, in panels (a) and (b), we find that the polelike peaks of Im$\S^\text{nor}$ (blue curve) and of Im$W$ (green) cancel out and leave no trace in their sum (red) at the peak energy.
As $T$ lowers from $0.12$ to 0.10 [panels (a) to (b)], the polelike peaks diminish and eventually below $T=0.06$, Im$W$ becomes negative everywhere so that no cancellation can occur anymore [(c) and (d)].
The cancellation is not found of course in the BCS region, where no prominent peak in Im$\S^\text{nor}$ is present.

Turning back to Figs.~\ref{dos}(c) and (d), we see that the appearance of the low-energy pole makes $\S^\text{ano}(\w =0)=\text{Re}\S^\text{ano}(0)$ depart from the static value, $\S^\text{ano}(\infty)=\text{Re}\S^\text{ano}(\infty)=U\langle a_{i\ua}a_{i\da}\rangle$.
We utilize the difference,
$\varDelta\S^\text{ano}\equiv \S^\text{ano}(0)-\S^\text{ano}(\infty)$,
to measure the significance of the self-energy pole on low-energy physics.
Figure \ref{pd}(b) plots $\varDelta\S^\text{ano}$ as a function of $T$ for various values of $U$.
In contrast to the absence of any appreciable $T$ dependence at weak couplings ($|U|<4$),  pronounced negative peaks appear at strong couplings ($|U|>4$) just below $T_\text{c}$. 
In Fig.~\ref{pd}(c), we have superposed the intensity map of $-\varDelta\S^\text{ano}$ on the phase diagram.
The strong intensity is found only on the fringe of the strong-coupling superconducting phase.
This colored ``dynamical" area is characterized by the presence of the self-energy poles with the canceling property, as discussed above.

%%%%%%%%%%%%%%%%%%%%%%%%%%%%%%%%%%%%%%%%%%%%%%%%%%%%%%%%%%%%%%%%%%%%%%%%%%%%%%%%%%%%%%%%%%%%%%%%%%%%%%%%%

%---hidden fermion
As it is shown in Ref.~\onlinecite{sakai14}, the cancellation of poles suggests a presence of a hidden {\it fermionic} excitation ($f$) hybridizing with a quasiparticle ($c$).
For sake of clarity, let us briefly repeat a proof of this statement.
The hybridization $V$ between $c$ and $f$ would be most simply accommodated in the phenomenological Hamiltonian,
\begin{align}
H_\text{TCFM}&=\sum_\s \left[ \e_c c_{\s}^\dagger c_{\s} + \e_f f_{\s}^\dagger f_{\s} + V (c_{\s}^\dagger f_{\s} + f_{\s}^\dagger c_{\s} ) \right] \nonumber\\
&-(D_c c_{\ua}c_{\da} + D_f f_{\ua}f_{\da} + \text{h. c.}),
\label{TCFM}
\end{align}
where $\e_c (\e_f)$ is the bare one-particle energy of the $c (f)$ fermion. 
For non-zero values of $D_c$ and $D_f$, the system is in the superconducting state.
By integrating out the $f$ degree of freedom in the path integral of the corresponding action, we obtain the normal and anomalous self-energies of the $c$ fermion as
\begin{align}
\S_c^\text{nor}({\w})&=V^2\frac{\w+\e_f}{\w^2-\e_f^2-D_f^2},\label{Snor}\\
\S_c^\text{ano}({\w})&=D_c-V^2\frac{D_f}{\w^2-\e_f^2-D_f^2}.\label{Sano}
\end{align}
These equations show that both self-energies have poles at the same energies, $\w=\pm\sqrt{\e_f^2+D_f^2}$, consistently with Figs.~\ref{dos}(b) and (d).
By defining $W_c(\w)=\frac{\S_c^\text{ano}(\w)^2}{\w+\e_c+\S_c^\text{nor}(-\w)^\ast}$ similarly to Eq.~(\ref{W}), we can calculate the residues of the poles at $\w=\pm\sqrt{\e_f^2+D_f^2}$ for both $\S_c^\text{nor}$ and $W_c$, to obtain $\frac{V^2}{2}\left(1\pm \frac{\e_f}{\sqrt{\e_f^2+D_f^2}} \right)$ and $-\frac{V^2}{2}\left(1\pm \frac{\e_f}{\sqrt{\e_f^2+D_f^2}} \right)$, respectively. Therefore, their residues cancel out in their sum.
In Supplementary Information, we show that Eqs.~(\ref{Snor}) and (\ref{Sano}) indeed fit the DMFT self-energies nearly perfectly at low frequencies when the cancellation occurs [Figs.~\ref{cancel}(a) and (b)]. On the other hand, Eqs.~(\ref{Snor}) and (\ref{Sano}) are not compatible with Figs.~\ref{cancel}(c) and (d) where no cancellation is observed. The latter is better fitted by a fermion-boson model \cite{sakai14}.
Thus, the pole cancellation gives a stringent test for a fermionic origin of the self-energy peaks.

Interestingly, the model (\ref{TCFM}) describes well the low-energy DMFT results in the normal state, too. 
In fact, the single polelike structure of $\S^\text{nor}$, seen for $T=0.13$ in Fig.~\ref{dos}(b), is consistent with Eq.~(\ref{Snor}), which yields
\begin{align}
\S_c^\text{nor}(\w)=\frac{V^2}{\w-\e_f} \label{Scnor}
\end{align}
for $D_c=D_f=0$. 

%---atomic limit
This observation gives us a clue to identify the hidden fermion $f$ in Eq.~(\ref{TCFM}).
In the strong-coupling region close to the atomic limit, the normal-state Green's function is well approximated by
\begin{align}
G_\text{atm}(\w)&=\frac{1-m}{\w+\mu}+\frac{m}{\w+\mu-U}\nonumber\\
   &=\left[ \w+\mu-Um-\frac{U^2m(1-m)}{\w+\mu-U(1-m)}\right]^{-1}\label{Gatm}
\end{align}
with $m$ being the average $c$-fermion density per spin.
Then, comparing Eq.~(\ref{Scnor}) with the second line of Eq.~(\ref{Gatm}), we find that $G_\text{atm}$ is reproduced by the model (\ref{TCFM}) with $V=U\sqrt{m(1-m)}$ and $\e_f=U(1-m)-\mu$.

Thus, the hidden fermion in the normal state, generating the pseudogap, is identified with 
the hidden fermion that appears by taking the atomic limit of the Hamiltonian (\ref{HM}).
It is then remarkable that the DMFT self-energy changes continuously across $T_\text{c}$ [Figs.~\ref{dos}(e) and (f)] since the continuity suggests that $f$ found below $T_\text{c}$ is a smooth continuation of the one originating the pseudogap above $T_\text{c}$ (where the atomic limit is most valid).
The local origin of $f$ in turn indicates its ubiquitous nature in the strong-coupling region, irrespective of dimensionality and lattice structures.

%---Fermionic area

\begin{figure}[tb]
\center{
\includegraphics[width=0.48\textwidth]{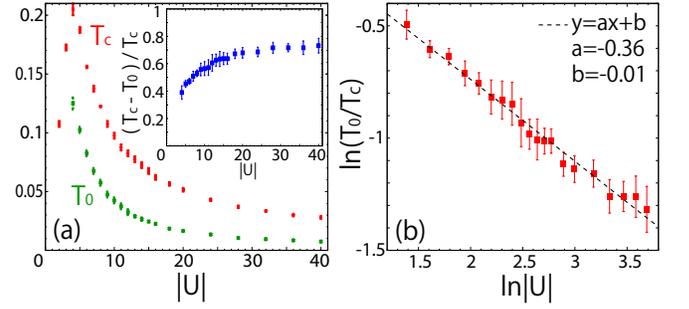}}
\caption{(Color online) (a) $T_\text{c}$ and $T_0$ (defined in the text) plotted against $|U|$. Inset shows a ratio, $(T_\text{c}-T_0)/T_\text{c}$, of the fermionic region to the superconducting region. (b) Scaling of $T_0/T_\text{c}$ to $|U|$.}
\label{T0}
\end{figure}

Thus, we have revealed the presence of the hidden {\it fermionic} excitation which dominates the quasiparticle dynamics in the strong-coupling superconductor just below $T_\text{c}$.
In order to quantify the area involving the hidden fermion, we define $T_0$ as a temperature at which $\varDelta\S^\text{ano}$ changes its sign from positive (for $T<T_0$) to negative (for $T>T_0$).
Below $T_0$, the order parameter is almost saturated [Fig.~\ref{pd}(a)] and the self-energy becomes nearly frequency independent [Fig.~\ref{dos}(c)]. 

Figure \ref{T0}(a) and the inset show that the hidden fermion indeed takes a major part of the superconducting phase for strong couplings.
Furthermore, Fig.~\ref{T0}(b) shows a remarkable scaling, $T_0/T_c \sim |U|^{a}$ with $a \sim -0.36$, which indicates that the hidden-fermionic area covers all the superconducting phase in the strong-coupling limit.

%---comparison to U>0 
Lastly, we compare the above results with those obtained in the two-dimensional repulsive Hubbard model. 
In the latter, cluster extensions of the DMFT \cite{maier05,kotliar01} have ascribed the origin of the pseudogap 
and the associated low-energy pole in $\S^\text{nor}$ above $T_\text{c}$ to ``Mottness" emerging in the proximity to the Mott insulator \cite{maier02,civelli05,kyung06-2,stanescu06,liebsch09,sakai09,gull09,sakai10,lin10,sordi11,sakai13}. In this case too a two-component fermion model well describes the pseudogap above $T_\text{c}$ as the hybridization gap formed with a hidden fermion \cite{imada11}.
Below $T_\text{c}$, particle-hole symmetric poles appear in $\S^\text{nor}$ and $\S^\text{ano}$ \cite{haule07} in a fashion similar to the attractive case, displaying the same pole cancellation and a similar mechanism of opening the superconducting gap at zero of Re$[\w+\e-\mu+\S^\text{nor}(-\w)^*]$ \cite{sakai14}. 
On the other hand, an essential difference is in the energy scale of the pseudogap, which is of the order of $|U|$ in the attractive case while it is less than $t$ in the repulsive case, suggesting different characters of the hidden fermion $f$.
Another important difference is in the role of $f$ in the superconductivity:
In the repulsive case $f$ helps to enhance $T_\text{c}$ \cite{maier08,sakai14}, while in the attractive case $f$ competes with superconductivity (the gap function is slightly suppressed [Fig.~S2(a)], even though $\S^\text{ano}$ is enhanced by the hidden fermion [Fig.~\ref{dos}(c)]), disappearing at low temperatures where a larger superconducting gap is preferred.

%conclusion
To summarize, we find a hidden {\it fermionic} excitation on the fringe of the strong-coupling superconducting phase of the attractive Hubbard model.
Although the superconducting transition in the strong-coupling region can be regarded as a boson condensation of tightly-bound Cooper pairs \cite{micnas90}, our results give an alternative fermionic point of view: 
The quasiparticles hybridize with a hidden fermionic excitation, which originates from a strong-coupling effect, above $T_\text{c}$ and down to $\sim T_0$ below $T_\text{c}$.
Above $T_\text{c}$, the hidden fermion yields the pseudogap in the quasiparticle spectra.
It persists in the superconducting state down to $T\sim T_0$, where the atomic physics still dominates the quasiparticle dynamics.
It disappears below $T_0$, where the smooth crossover to a static superconducting state takes place. 
The hidden fermion region dominates for $|U|/D \rightarrow \infty$.
Many issues remain open: What is the physical entity and the explicit expression of $f$? 
Why does $\Sigma^{\rm ano}$ lose its frequency dependence below $T_0$?
Why does $T_0/T_\text{c}$ follow the scaling in Fig.~\ref{T0}(b)?
The presence of the hidden fermions in two different unconventional superconducting states (for $U>0$ and $U<0$) calls for future search of similar excitations in other unconventional correlated superconductors \cite{capone02-2,nomura15,han04,sakai04,hoshino14}.

\begin{acknowledgments}
S.S. acknowledges useful comments by G. Sangiovanni. The work was supported by JSPS KAKENHI Grant No. 26800179, the Computational Materials Science Initiative (CMSI), HPCI Strategic
Programs for Innovative Research (SPIRE), and RIKEN Advanced Institute for
Computational Science (AICS) (Grant No. hp130007, hp140215, hp150211) from
MEXT, Japan.
\end{acknowledgments}

\section{Supplementary Information}

\subsection{Method}

We solve the DMFT impurity problem with the exact diagonalization method with  8 bath sites for all the presented data while we have checked its validity with several calculations with 10 and 12 bath sites.

In order to correctly capture a spectral structure in the gapped states, we first arrange the bath sites in a one-dimensional-chain form at $T=0$ \cite{si94}. 
For calculations at $T>0$, we rearrange the bath sites so that the impurity site couples to every bath sites \cite{caffarel94}, by fitting the hybridization function (obtained by the above chain algorithm) on real-frequency axis.
This rearrangement offers an initial guess of the bath parameters at low temperatures.
We then apply the standard finite-$T$ ED algorithm of Ref.~62, with gradually increasing the temperature from $T=0$.
In the algorithm, we optimize the bath parameters at each temperature by minimizing the distance function, $d=\sum_n \left[ |g_0^\text{nor}(i\w_n)-\tilde{g}_0^\text{nor}(i\w_n)|+|g_0^\text{ano}(i\w_n)-\tilde{g}_0^\text{ano}(i\w_n)|\right]$ to fit the normal/anomalous component of the Weiss function ($g_0^\text{nor/ano}$) with that of 8 bath sites ($\tilde{g}_0^\text{nor/ano}$). 
Here $\w_n=(2n+1)\pi T$ is the Matsubara frequency.
Note that, while the above procedure improves the spectral structure beyond the gap energy, the low-energy structure of the self-energy (within the spectral gap) is robust against different choices of the initial guess.

To display the real-frequency properties, we introduce the energy-broadening factor, $i\h(\w)=i\h_0 \min[1+\w^2,10]$ to $\w$. We set $\h_0=0.05$ unless otherwise mentioned. The presented results do not essentially depend on the choice of $\h(\w)$.

\subsection{Superconducting gap function}

\begin{figure}[b]
\center{
\includegraphics[width=0.48\textwidth]{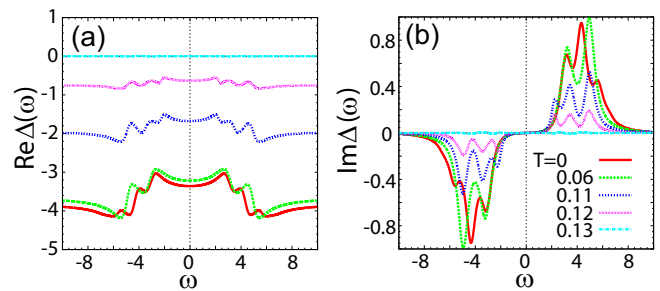}
}
\caption{The superconducting gap function at $U=-8$ for the same temperatures as those used in Figs.~2(a-d) in the main text.}
\label{Delta}
\end{figure}

Figure S\ref{Delta} shows the superconducting gap function, which is defined through $\S^\text{nor}$ and $\S^\text{ano}$ by \cite{scalapino66}
\begin{align}
\D(\w)&\equiv \frac{ \S^\text{ano}(\w) }{1-\frac{1}{2\w}\{\S^\text{nor}(\w)-\S^\text{nor}(-\w)^\ast\}}.\label{D}
\end{align} 
Interestingly, the $\w$-dependent structure of $\D(\w)$ is considerably different from that of $\S^\text{ano}(\w)$ [Figs.~2 (c) and (d) in the main text], suggesting a substantial role of the $\w$ dependence in $\S^\text{nor}(\w)$. 
The deviation manifests the departure from the BCS and Migdal-Eliashberg theories, which neglect the $\w$ dependence of $\S^\text{nor}(\w)$.

In particular, at $T=0.12$ where the peak of the hidden fermionic excitation is seen at $\w=\pm 0.9$ in Im$\S^\text{ano}$, Im$\Delta$ does not show any structure at the same energies.
Instead, Im$\Delta$ shows an intensity at higher energies ($|\w|\gtrsim 1.5$).

At $T=0$ and 0.06, Im$\Delta$ shows finite amplitudes only in a frequency range similar to that of Im$\S^\text{ano}$ while a simple linear relation between them (as assumed in the Migdal-Eliashberg theory) does not yet hold [see inset to Fig.~2(d) for comparison] as is apparent also from the residual $\w$ dependence in $\S^\text{nor}$ [inset to Fig.~2(b)].

\subsection{Continuous development of $\S^\text{ano}$ around $T_\text{c}$}
\begin{figure}[tb]
\center{
\includegraphics[width=0.48\textwidth]{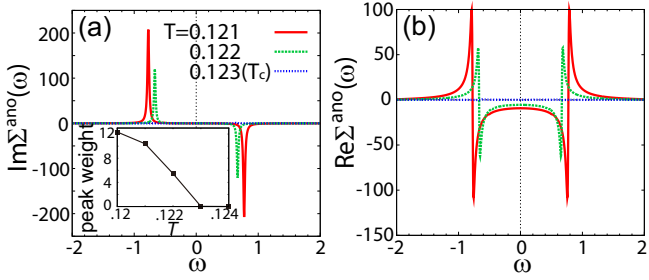}}
\caption{Change of $\S^\text{ano}$ just below $T_\text{c}\simeq 0.123$. Inset in panel (a) plots the weight of the low-energy peak of Im$\S^\text{ano}$ against $T$.
$\h_0=0.01$ is used to sharpen the pole and zero features. }
\label{aroundTc2}
\end{figure}

Figure S\ref{aroundTc2} shows the continuous development of $\S^\text{ano}$ around $T_\text{c}$, in relation to Figs.~2(e) and (f) in the main text.
The continuity would be most directly seen in the inset to panel (a), which plots the weight of the low-energy peak in Im$\S^\text{ano}$ against $T$. 
Despite this continuous change of $\S^\text{ano}$,  the singularity responsible to the spectral gap is replaced from a pole of $\S^\text{nor}$ (at $\w=\w_\text{nor}$) to a zero of Re$[\w+\e-\mu+\S^\text{nor}(-\w)^*]$ (just above $\w_\text{nor}$) immediately below $T_\text{c}$ as the cancellation of the self-energy poles occurs [Fig.~2(f)].
This is possible because the pole of $W$ (at $\w=\w_\text{nor}$) aquires a finite residue immediately below $T_\text{c}$, where the new zero of Re$[\w+\e-\mu+\S^\text{nor}(-\w)^*]$ is created.

\subsection{Fitting with the two-component fermion model}

\begin{figure}[tb]
\center{
\includegraphics[width=0.48\textwidth]{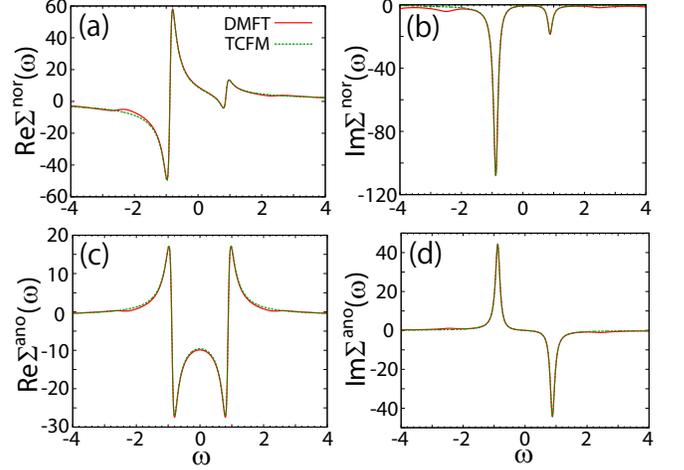}
}
\caption{Self-energies calculated with the DMFT (red curves) for $U=-8$ and $T=0.12$ and with the two-component fermion model (TCFM, green curves) for $\e_f=-0.630$, $V=3.39$, $D_f=-0.624$, and $D_c=U\langle c_\uparrow c_\downarrow \rangle =-0.736$.}
\label{fit}
\end{figure}

Figure S\ref{fit} shows the fitting of the DMFT self-energies with Eqs.~(5) and (6), with $D_c=U\langle c_\uparrow c_\downarrow \rangle =-0.736$ calculated by the DMFT.
We can see that the fitting works nearly perfectly for the low-energy peaks at $\w=\pm 0.9$; Only the visible deviation in the presented scale is the weak structures around $\w=\pm 2.5$ in the DMFT self-energies, which cannot be captured by the model (4).


\begin{thebibliography}{99}
\bibitem{note1}
Though not shown here, we have obtained qualitatively similar results for other dopings, too.
\bibitem{note3}
The multiple peaks (broadened by $\h$) in the DOS at $T=0$ is consistent with the previous works \cite{sangiovanni06,strack92} for the strong-coupling repulsive Hubbard model.
\bibitem{note5}
Here we have considered the momentum with $\e=0$ while the following discussion does not essentially depend on the choice of $\e$ since $\e\ (|\e|< 2)$ is negligible compared to $\S^\text{nor}$ in Eq.~(\ref{W}).
\end{thebibliography}
\end{document}